# Legal Aspects of Decentralized and Platform-Driven Economies

**Marcelo Corrales Compagnucci, Toshiyuki Kono and Shinto Teramoto**

**Abstract** The sharing economy is sprawling across almost every sector and activity around the world. About a decade ago, there were only a handful of platform-driven companies operating on the market. Zipcar, BlaBlaCar and Couchsurfing among them. Then Airbnb and Uber revolutionized the transportation and hospitality industries with a presence in virtually every major city. "Access over ownership" is the paradigm shift from the traditional business model that grants individuals the use of products or services without the necessity of buying them. Digital platforms, data and algorithm-driven companies as well as decentralized blockchain technologies have tremendous potential. But they are also changing the "rules of the game." One of such technologies challenging the legal system are AI systems that will also reshape the current legal framework concerning the liability of operators, users and manufacturers. Therefore, this introductory chapter deals with explaining and describing the legal issues of some of these disruptive technologies. The chapter argues for a more forward-thinking and flexible regulatory structure.

**Keywords** sharing economy, platforms, AI, blockchain, data protection, autonomous vehicles

# 1 Introduction

The sharing economy is a new model of organizing economic activity that may substitute traditional corporations and capitalism around the world. This activity is based on acquiring, providing or sharing access to goods and services based on underutilized assets.[1] The sharing economy is facilitated by a community based on digital platforms that enable people who have never met before to share resources and trust each other. Information and innovation technologies are used in order to match individuals who possess such superfluous resources with existing demand in the market.[2]

Decentralized platforms are sometimes associated with "peer-to-peer" (P2P) technologies (such as Napster in the late 1990s). However, it is characterized by two very different business models: i) Sometimes *individuals* own and get to share their assets with each other, like extra rooms (Airbnb[3] and Homestay),[4] cars (Uber,[5] Lyft and Turo),[6] parking spaces (Just Park),[7] skills (Taskrabitt)[8] and even their own pets (Rover);[9] ii) In other cases, *companies* own and lend out the assets, such as cars, bicycles and motorbikes. Prime examples are Zipcar[10] and Car2Go[11] now owned by Avis and Daimler-Benz, respectively.[12]

It is indisputable that the sharing economy produces an enormous amount of wealth. In 2015, PriceWaterhouseCoopers projected growth from $15 billion in global revenue to $335 billion in 2025.[13] Faster than ever, we have barely begun to scratch the surface of the possibilities of innovation and dynamic capability behind this new economic model.[14] However, the extraordinary growth of the sharing economy creates unprecedented legal problems. This book anthology takes up various disruptive technologies that are currently transforming the legal system around the world. Such technologies include: cloud computing, Big Data, Internet of Things (IoT), artificial intelligence (AI), machine learning (ML), deep learning (DL), blockchain, algorithms and other related autonomous systems – such as self-driving vehicles.

There is no doubt that the expansion of the sharing economy is changing our world. It has also triggered the emergence of new products and services. As a result, companies are becoming increasingly more data and algorithm-driven, making use of so-called "decentralized platforms." New transaction and/or payment methods such as Bitcoin, Ethereum, etc., which are based on trust building systems using blockchain, smart contracts and other distributed ledger technology (DLT) also constitute an essential part of such a new economic model and central to the analysis of this work.

---

[1] Sundararajan (2016).
[2] See, e.g., Malik and Wahaj (2019), pp. 249 et seq.
[3] Maurer (2016), p. 6.
[4] See Homestay. Available at: https://www.homestay.com. Accessed 10 June 2019.
[5] See Uber. Available at: https://www.uber.com. Accessed 10 June 2019.
[6] See Turo. Available at: http://www.turo.com. Accessed 10 June 2019.
[7] See Justpark. Available at: https://www.justpark.com. Accessed 10 June 2019.
[8] See Taskrabbit. Available at: https://www.taskrabbit.com. Accessed 10 June 2019.
[9] See Rover. Available at: https://www.rover.com. Accessed 10 June 2019.
[10] See Zipcar. Available at: https://www.zipcar.com. Accessed 10 June 2019.
[11] See Car2go. Available at: https://www.car2go.com/US/en/. Accessed 10 June 2019.
[12] Cusumano (2018), pp. 26-28.
[13] PriceWaterhouseCooper (2015), pp. 1-30.
[14] Gazola (2017), pp. 75-94.

Most of these digital platforms within the sharing economy rely on cloud-based infrastructures to operate at the upper level. This paradigm shift would not be possible without the adoption of cloud computing deployment models and services. Individuals and companies in general, are fast gearing up for the "on-demand" and "pay-as-you-go" culture, which constitute the building blocks of cloud computing transactions.[15]

All these new technological breakthroughs have brought complex ways of processing and analyzing information at a larger scale.[16] From a legal perspective, the uncertainties triggered by the emergence of a new digital reality are particularly urgent. How should these tendencies be reflected in legal systems in each jurisdiction? This collection brings together a series of contributions by leading scholars in the emerging field of the sharing economy. The aim of this book is to enrich legal debates on the social, economic and political meaning of this new economic model along with these cutting-edge technologies.

It is indisputable that this paradigm shift is changing the scope in which law is designed, interpreted and applied in a constantly evolving environment. There is, therefore, an increasing awareness that the traditional concepts and approaches of the law must be more flexible and expanded to encompass new areas associated to this new economic model. Based on this new reality, this work aims to provide insights on some of the key legal topics that will affect the future of our daily lives. The chapters presented in this edition attempt to answer some of these questions from the perspective of different legal backgrounds. The aim is to answer some of these questions from an inter-disciplinary and integrated point of view taking into account a variety of legal systems.

## 2 Parts

Addressing the many challenges created by the sharing economy requires going beyond one single disciplinary perspective or frame of reference. As such, after this introductory chapter, the book is divided into four parts comprising 12 chapters as follows: Part I – "Sharing Economy & Platforms;" Part II – "Digital Age and Personal Data;" Part III – "Blockchain & Code;" and, Part IV – "Autonomous Systems & Future Challenges." Each part focuses on one particular area of the sharing economy by adopting different approaches and methods.

Part I – "Sharing Economy & Platforms" – focuses on the impact of the sharing economy as an emerging economic model. The chapters in this section cut across different aspects of digital platforms, including platforms for building collaboration and the governance of cities in real world settings using the use case examples of sharing cities in Seoul and the study of the Japanese housing accommodation legal system.

---

[15] Tandon (2018).
[16] See, e.g., Chen et al. (2014), pp. 12 et seq.

Part II – "Digital Age & Personal Data" – has the digital person and the protection of personal data as its overarching subject. The starting point is that there is the need to discuss the concept of "digital person" as a new legal entity in light of the development and application of AI technology. This takes us directly to the necessity of changing the legal education system and the role of future lawyers as this will affect the design of the new global architecture. This part also focuses on the protection of the personal data of individuals in the context of health systems. It does this by comparing different approaches of consent under the EU General Data Protection Regulation (GDPR)[17] and the specific situations in Australia and the United Kingdom (UK). The GDPR has been generally well received for strengthening some of the rules in the previous EU Data Protection Directive[18] by granting individuals more control over their data when using electronic health records. However, it has also generated hot debate around the world regarding its practicability and flexibility within modern processing technologies.

Part III – "Blockchain & Code" – discuss blockchain technologies and how learning the legal issues embedded in the "code" of computer software can help legal professionals to reinvent themselves. Some think that lawyers or even bankers and notaries might soon have to offer new services and change the way they do business otherwise they might become redundant and obsolete.[19] The blockchain is one of the most hyped terms of this new century and it has been said that it can revolutionize the world.[20] Overall, the chapters of this part discuss some of the main features of blockchain technology to increase trust and transparency in decentralized networks.

Part IV – "Autonomous Systems & Future Challenges" – looks into some of the legal implications of autonomous vehicles. Autonomous driving represents a crucial part of the mobility of the future. However, there are still vexed legal issues. What happens for example when there is a car accident? Who should be liable for it? Is it the AI or automated system software developer? Or, is it the auto manufacturer who assembled the car pieces together? How shall we handle the insurance companies? In sum, how can we reduce and mitigate these legal risks?[21] These are just some of the main questions that come to our minds immediately when we are talking about driverless cars. The chapters in this section explore some of the legal problems and suggest that this will affect the sharing economy, which might not be ready yet to face these problems. Finally, the remaining chapter of this part discusses the protection of trade secrets in light of the new EU Directive on the protection of trade secrets and outlines the options for implementation.

---

[17] Regulation (EU) 2016/679 of the European Parliament and of the Council of 27 April 2016 on the protection of natural persons with regard to the processing of personal data and on the free movement of such data, and repealing Directive 95/46/EC (General Data Protection Regulation). While the Regulation entered into force on 24 May 2016, it applies to all EU Member States from 25 May 2018. See European Commission, Reform of EU Data Protection Rules https://eugdpr.org. Accessed 10 June 2019.
[18] Directive 95/46/EC of the European Parliament and of the Council of 24 October 1995 on the protection of individuals with regard to the processing of personal data and on the free movement of such data.
[19] Vermeulen (2017).
[20] Corrales, Fenwick and Haapio (2019), pp. 2 et seq.
[21] https://www.daimler.com/innovation/case/autonomous/legal-framework.html

# 3 Chapters

After this introduction, the book comprises twelve substantive chapters. Part I – *Sharing Economy & Platforms* – consists of three contributions.

*Annelise Riles* starts by explaining the collaborative genius of today's interconnected world. The image of the lonely genius working by himself in his office is largely over. The pervasive and dynamic nature of our current society presents difficult technical and legal questions. Transactions frequently take part in complex relationships where several actors are involved across different jurisdictions. Therefore, the genius of our time is a collaborative genius. From the business world to the academy, and from the leading financial centers to grassroots development projects around the world, collaboration is increasingly perceived as a necessity. This chapter deals with some of these challenges, and also some of the possibilities that are inherent in collaboration, taking into account the example of a recent experiment with Meridian 180, a global engagement platform for policy experimentation founded in 2011.

*Benjamen Franklen Gussen* retains a focus on the impact of the sharing economy on the governance of cities using the lessons learnt from the Sharing City Seoul Project. The chapter starts by explaining the historical grounds of collaboration among societies. The analysis suggests that collaboration was always present in different forms from the beginning of civilization. The nature of collaboration, however, shifted to a hierarchical organizational model as a result of increasing population density. This led to the emergence of new thriving cities and markets. The arrival of new waves of technological innovation in the 21$^{st}$ century, however, brought with it a new change to a network organization and collaboration. This innovation allows for a return to a network organization at a scale (population-density) never seen before. Seoul is a very good example of this paradigm shift as it is one of the leading cities when it comes to innovation in the sharing economy. Its experience has influenced already other major cities across the world. The chapter explains the policy interventions that took place in Seoul to inform future approaches to governing cities in other countries and looks at the role of law in enabling cities as a supple network of digital technologies that stimulate innovation within the sharing economy.

*Yuichiro Watanabe* focuses on a study of the Japanese Housing Accommodation Business Act (Act No. 65 of 2017), which is the first national law in the world legitimizing home-sharing. The starting point of this chapter is to explain the making process of the new Japanese Housing Accommodation Business Act, which came into effect in 2018 after 3 years of negotiations. The chapter also explains the previous regulatory framework which had remained almost untouched since 1948. The chapter identifies and reviews three legal issues: (a) extraterritorial application outside of Japan; (b) lacking the consistency between the other existing Japanese laws in terms of regulating digital platforms; and (c) illegal local ordinance beyond the Act, preventing the sharing economy by imposing additional restrictions. Finally, the author suggests that the Act might not be the best model to be implemented in other jurisdictions.

Part II – *Digital Age & Personal Data* – comprises three chapters.

*Cecilia Magnusson Sjöberg* discusses the concept of "digital person" as a new legal entity taking into consideration the recent development and application of AI technology. The purpose is not to just add another term to those of the natural person and the legal person, but to introduce a concept that could eventually, under certain circumstances, be implemented in the legal order. This new legal figure could also, tentatively, be granted legal capacity, with rights and responsibilities. This new legal person i.e., the "digital person" would as such be possible to describe as a constellation of algorithms consisting of a basic algorithmic identity, which could be profiled and specified with reference to various purposes. The risk is otherwise an emerging dysfunctional legal society where there is no legal entity, which can take the role of the subject. Issues that arise concern e.g., self-driving car liability, pricing algorithms on the competitive market and data protection when profiling consumers.

In their chapter, *Janos Mészáros, Chih-Hsing Ho and Marcelo Corrales Compagnucci* examine the challenges of the revised opt-out system and the secondary use of health data in the UK. The secondary use of data refers to the processing of data collected during direct care for new purposes, such as research and policy planning. The analysis of this data could be very valuable for science and medical treatment as well as the discovery of new drugs. For this reason, the UK government established the "care.data program" in 2013. The aim of the project was to build a central nationwide database for research and policy planning. However, the processing of personal data has been planned without proper public engagement, and the central database was aimed to be used by public and private third-parties, including IT companies. The care.data program established a double opt-out system which turned out to be very controversial due to scandals such as the Google DeepMind deal with the U.K.'s National Health Service (NHS).[22] Google's artificial intelligence firm was allowed to access health data from over 1.6 million patients to develop an app monitoring kidney disease called "Streams." Public concerns and corroborative research studies suggested that the Google DeepMind deal had access to other kinds of sensitive data and failed to comply with data protection law.

For this reason, since May 2018, the UK government launched the "national data opt-out" (ND opt-out) system[23] with the hope to regain the public trust. Nevertheless, there is no evidence of significant changes in the ND opt-out. Neither in the use of secondary data, nor in the choices that patients can make. The only notorious difference seems to be in the way that these options are communicated and framed to the patients. Most importantly, according to the new ND opt-out, the type-1 opt-out option – which is the only choice that truly stops data from being shared outside direct care – will be removed in 2020. According to the Behavioral Law and Economics literature (Nudge Theory),[24] default rules – such as the revised opt-out system in the UK – are very powerful, because people tend to stick to the default choice made readily available to them. The crucial question analyzed in this chapter is whether it is desirable for the UK government to stop promoting the type-1 opt-outs, and whether this could be seen as a kind of "hard paternalism."

---

[22] See National Health Service (NHS). Available at: https://www.nhs.uk. Accessed 10 June 2019.
[23] See "national data opt-out" (ND opt-out) system Available at: https://digital.nhs.uk/national-data-opt-out. Accessed 10 June 2019.

*Danuta Mendelson* focuses on the National Electronic Health Record System and Consent to the Processing of Health Data in the European Union (EU) and Australia. She does this by comparing the legal framework in the EU within the scope of the GDPR in the context of the EU national electronic record (NEHR) schemes with the approach of the Australian national health record scheme called My Health Record (MHR).[25] The main difference being the different levels of developments of consent between the two approaches. In the EU, the GDPR proclaims that individuals (data subjects) should give their consent for the processing of their personal health data or have the right to refuse the processing of their personal health data not just in general, but in specific instances. Whereas in Australia, legislators did not predicate the lawfulness of personal health data processing on the individual data-subject's valid and informed consent. For this reason, the creators of the MHR system did not embed in its architecture the right of patients to give free, informed, specific and unambiguous indication that they agree to the processing of their personal health data in every (reasonable) instance. This might be one of reasons why Australia is not seeking certification for attainment of adequate level of data protection from the European Commission. The chapter concludes that under the MHR scheme in Australia, data subjects – in particular those vulnerable patients such as the very young and the elderly – will remain disempowered data subjects.

Part III of the book – *Blockchain & Code* – contains three contributions.

In their chapter, *Mark Fenwick, Wulf Kaal and Erik Vermeulen* discuss the importance for lawyers and law students to learn how to code in this new digital age. The authors explain the benefits of introducing a Coding for Lawyers course in the legal curriculum and they share their initial experiences with the course. The main argument is that Legal Technology (LegalTech) is profoundly disrupting the legal profession and the development of cutting-edge technologies – such as blockchain, AI, Big Data, smart contracts, etc. – have triggered the emergence of new business models. The authors conclude that since all these technologies are code-based, lawyers – as active "transaction engineers" – need to be able to understand and communicate in and about code to participate in the design of such technologies. The computer code is ubiquitous by nature, yet hidden. This is affecting the legal framework, particularly in terms of the on-going legal profession. Therefore, lawyers of the future will be transaction engineers managing the responsible deployment of new technologies and the design of a new global architecture, and that to perform

---

[24] See, e.g., Thaler and Sunstein (2009).
[25] See My Health Record. Available at: https://www.myhealthrecord.gov.au. Accessed 10 June 2019.

Electronic copy available at: https://ssrn.com/abstract=4027412

this function effectively, legal professionals need to develop a number of new skills and capacities, including an understanding of the basic concepts and power of coding.

*Craig Calcaterra and Wulf Kaal* explain the reputation protocol for the Internet of Trust. Trust is very important and became one of the most important drivers in our global economy. Internet-based platform business outcompetes traditional business and DLT shows a lot of promise in multiple business vertical. Studies suggest, however, that they have not reached their full potential due to the decreasing global trust in the Internet and under-developed trust in decentralized technology solutions. Semada and the Semada Research Institute (SRI)[26] believe that a decentralized reputation network can reverse that trend to increase trust in the Internet and increase decentralized technology adoption. The authors call this solution the Semada Internet of Trust – a network that uniquely captures real world information, context, and value in cryptographic transactions generating transparently validated consensus on truth. Through the creation of unconscious and conscious trust in decentralized network adoption becomes a desirable outcome and increases. The Semada Internet of Trust creates a framework of trust through reputation and incentive optimization that enables market conditions for unprecedented business models and market capitalization while reducing adverse selection and moral hazards for transacting parties. The network will provide information symmetry while reducing transaction costs for market participants. Businesses will leverage the platform to achieve efficiencies across verticals.

*Gyooho Lee* discusses two issues raised by the Korean legal community in terms of blockchain technology. One is intellectual property-related issues of open source software on which blockchain technology is based. When it comes to open source software, one pivotal court case needs to be explored. It is related to the conflict between the possessor of a trade secret and copyright owner of open source software. The other issue concerns how to guarantee the authenticity of e-Apostilles by using blockchain technology. The lack of authenticity of e-Apostilles is a big legal hurdle which prevents e-Apostilles from gaining popularity in many countries. Blockchain computing is a good solution for guaranteeing the authenticity of e-Apostilles. The chapter explores an IP-related issue which inherently concerns blockchain technology and proposes a method for ensuring the authenticity of e-Apostilles by using blockchain technology. In sum, the chapter illustrates an inherent legal issue of blockchain technology and blockchain technology as a method to solve a current legal problem.

Part IV of the book – *Autonomous Systems & Future Challenges* – comprises three chapters.

*Shinto Teramoto* explains the issues of autonomous driving from a lawyer's perspective. The development of driverless cars is still in its early stages. At this point, the focus of attention should be on the safety and user-friendliness of autonomous driving on public roads. Therefore, the current discussion must be who can effectively prevent traffic incidents and accidents involving autonomous driving by investing their own financial, human, and technological resources. Safe autonomous driving is a product of a well-organized network system. Contending hastily that only specific classes of nodes should be liable does not make sense. The importance of telecommunication predicts that telecom and network service industries will play major roles in realizing safe road traffic involving autonomous driving. If lawyers discuss the allocation of cost caused by traffic incidents or accidents involving autonomous driving without considering the involvement of telecom and network service industries, it simply shows the carelessness of lawyers. Autonomous driving is the way to realize the sharing of surplus resources, which have been unused or wasted, by means of aggressive involvement of information and communication technology (ICT). The chapter concludes with a suggestion that lawyers should go beyond the current debate of liability issues caused by traffic accidents and incidents involving autonomous driving. Yet, the discussion should involve collaborative ways of developing interfaces and standards for every road traffic participant to effectively and efficiently communicate with each other.

---

[26] See Semada Research Institute (SRI). Available at: https://semada.io. Accessed 10 June 2019.

*Steven van Uytsel and Danilo Vasconcellos Vargas* discuss some of the legal implications of adversarial machine learning. Research revealed that perturbations to a picture – even in small size – may disable a deep neural network from correctly qualifying the content of a picture. This research has been transplanted to traffic signs. The test results were disastrous. For example, a perturbated stop sign was recognized as a speeding sign. Because visualization technology is not able to overcome this problem yet, the question arises who should be liable for accidents caused by this technology. Manufacturers are being pointed at and for that reason it has been claimed that the commercialization of autonomous vehicles may stall. Without autonomous vehicles, the sharing economy may not fully develop either. This chapter shows that there are alternatives for the unpredictable financial burden on the car manufacturers for accidents with autonomous cars. This chapter refers to operator liability, but argues that for reasons of fairness, this is not a viable choice. A more viable choice is a no-fault liability on the manufacturer, as this kind of scheme forces the car manufacturer to be careful but keeps the financial risk predicable. Another option is to be found outside law. Engineers could build infrastructure enabling automation. Such infrastructure may overcome the problems of the visualization technology, but could potentially create a complex web of product and service providers. Legislators should prevent that the victims of an accident, if it were still to occur, would face years in court with the various actors of this complex web in order to receive compensation.

The final chapter, by *Thomas Hoeren* explores some of the provisions enshrined in the new EU Directive on the protection of trade secrets (Directive (EU) 2016/943).[27] The Directive was adopted in June 2016 and is now to be transposed into national law by June 2018. This led to discussion at national levels whether the protection of trade secrets should be included in a comprehensive set of rules of intellectual property rights, or at least regulated by a special law. The chapter focuses on the producer's liability in accordance with Article 4 (5) of the Directive and outlines the options for implementation. The chapter concludes that the Directive will permanently change European secrecy law. Compared to current German regulations, important differences can be observed, concerning for example the definition of a trade secret or the legitimacy of reverse engineering. Therefore, the final implementation of the Directive can be suspenseful. In the meantime, companies are summoned to conduct concrete confidentiality measures and to adjust to the changed field of secrets. In this context, extended nondisclosure agreements and a strategy to deal with the freedom of reverse engineering are most important. A new culture of secrecy in companies and with suppliers must be added.

---

[27] Directive (EU) 2016/943 of the European Parliament and of the Council of 8 June 2016 on the protection of undisclosed know-how and business information (trade secrets) against their unlawful acquisition, use and disclosure.